 \documentstyle[epsfig,twocolumn,prb,aps]{revtex}
 \newcommand{\bm}[1]{\mbox{\boldmath ${#1}$}}
\begin{document}


\title{
Strongly reduced gap in the zigzag spin chain with a ferromagnetic 
interchain coupling
}

\author{
Chigak Itoi${}^{1}$ and Shaojin Qin${}^{2}$,
}
\address{
${}^1$ Department of Physics, Nihon University, Kanda, Surugadai,
 Chiyoda, Tokyo, Japan\\
${}^2$ Institute of Theoretical Physics, CAS, Beijing 100080,
 P R China\\
}

\date{\today}
\maketitle
\begin{abstract}
We study a spin 1/2 Heisenberg zigzag spin chain model near decoupled 
two chains. Taking into account a symmetry breaking perturbation, 
we discuss the existence of an energy gap in the 
ferromagnetic interchain coupling as well as the antiferromagnetic one.
In the ferromagnetic model, a marginally relevant fixed line reduces 
the gap strongly, so that the correlation length becomes an 
astronomical length scale even in order 1 coupling.  This result agrees 
with density matrix renormalization group results.

\end{abstract}


\section{Introduction}

It is interesting to study the effects of a frustration in a quantum system.
The zigzag spin chain is one of the simplest quantum spin models with a
frustration. An experiment on the compound ${\rm SrCuO_2}$  reports a 
remarkable suppression of the three dimensional ordering temperature 
due to a strong quantum fluctuation enhanced by the frustration 
\cite{MK}.  Here we study the effect of frustration in the zigzag chain 
with the following Hamiltonian 
\begin{equation}
H= \sum_i
\left[ 
J_2 ({\bf S}_{i} \cdot {\bf S}_{i+1} 
+ {\bf T}_i \cdot {\bf T}_{i+1}) 
+J_1 {\bf S}_i \cdot ({\bf T}_{i}+{\bf T}_{i+1})
\right] . \nonumber
\label{hamzg}
\end{equation}
The operators ${\bf S}_i$ and ${\bf T}_i$ are $s=1/2$ spins.  Here we 
treat an interchain coupling $J_1$ as a perturbation to the two 
decoupled antiferromagnetic ($J_2 > 0$) spin chains. 

It is well-known that the zigzag chain model with antiferromagnetic 
region $J_1 > 0$ has a dimer phase and gapless phase \cite{H}.  The 
region of the dimer phase \cite{N} $( 0 < J_1 < J_{c}  \sim 4.15 J_2)$ 
contains a Majumdar Ghosh point $J_1 = 2 J_2$ where the dimerized 
ground state is obtained exactly \cite{MG}.  The point $J_1=0$ is a 
critical point of the two decoupled antiferromagnetic chains. There is 
another critical point $J_1=-4 J_2$, where the level crossing between 
singlet and fully polarized ferromagnetic ground states occurs. There 
is an exact solution \cite{HN} of a fully polarized ferromagnetic 
ground state in $J_1 < -4 J_2$, and for a doubly degenerated ground 
state at the critical point $J_1 = -4 J_2$.  A numerical analysis in a 
region $J_1 > -4 J_2 $ indicates a complicated size dependence of the 
ground state energy \cite{TH,B}. 

It has been long believed that this model is gapless \cite{WA,AS} for a 
small ferromagnetic region $J_1 < 0$. In several papers however, one 
loop renormalization group (RG) shows an instability of the critical 
point $J_1=0$ both in ferromagnetic ($J_1 < 0$) and antiferromagnetic 
($J_1 > 0$) region due to a Lorentz symmetry breaking perturbation 
\cite{NGE,CHP}. This fact is a puzzling because the expected energy gap 
produced by the unstable flow has never been observed in the 
ferromagnetic region either numerically or experimentally. If it were 
gapless, there should be a new stable critical point missed in the one 
loop approximation.  In this paper, we clarify a natural mechanism to 
solve this problem. We conclude that the actual energy gap 
is finite but very tiny in an extended region of the 
coupling constant space. We can understand unstable the RG flow 
and numerical analysis consistently. The energy gap is
too small to observe in any numerical method. 

A stable large scale reduction 
is an important and difficult fundamental problem in theoretical physics.
In particle physics, any unified theory has this kind of hierarchy 
problem. Generally speaking, a unified theory 
has only one large energy scale, and also should explain 
the generation of a small energy scale to describe the low energy
physics. These two requirements  
makes unified theories quite unnatural, 
since we need a fine tuning of the coupling constants for large 
scale reduction. We believe that 
nature do not chose special value of the coupling constants.
From the view point of hierarchy problem in the theoretical physics, 
this zigzag chain model can be an interesting example in statistical physics.
We conclude that the very tiny gap is always
obtained in an extended region of the ferromagnetic interchain coupling
unlike the antiferromagnetic coupling. 

This paper is organized as follows. In section II we construct an
effective field theory for the zigzag chain and calculate the beta functions. 
In section III, we study the RG flow and discuss an
energy gap reduced strongly in the ferromagnetic interchain coupling region.
In section IV, we calculate the energy spectrum to describe the results
of numerical calculation. In section V, we show our numerical analysis. 

\section{Effective field theory}
The unperturbed theory is two decoupled Heisenberg antiferromagnetic 
chains whose effective theory is two decoupled $SU(2)_1$ WZW models 
\cite{NGE}.  We introduce two free bosons $\varphi_a(z,\bar{z})
=\varphi_a(z)+\bar{\varphi}_a(\bar{z})$ $(a=1, 2)$ with 
\begin{equation}
S_0 = \frac{1}{2 \pi} \int d^2 z
\left( \partial \varphi_1 \bar{\partial} \varphi_1
+\partial \varphi_2 \bar{\partial} \varphi_2 \right).
\end{equation}
with two point functions 
$\langle \varphi_a(z) \varphi_b(0) \rangle = - \delta_{ab} \log z$, 
$\langle \bar{\varphi}_a(\bar{z}) \bar{\varphi}_b(0) \rangle
 = - \delta_{ab} \log \bar{z},$ for two WZW models.
The spin operator is written in terms of two bosons
\begin{eqnarray}
2 \pi \ {\bf S}_j &=& {\bf J}_1 + \bar{\bf J}_1 
		+ (-1)^j M tr [(g_1 + g_1 ^{\dag})
			\frac{^t{\bm \sigma}}{2}], \\
2 \pi \ {\bf T}_j &=& {\bf J}_2 + \bar{\bf J}_2 
		+ (-1)^j M tr [(g_2 + g_2 ^{\dag})
			\frac{^t{\bm \sigma}}{2}],
\end{eqnarray}
where $M$ is a nonuniversal real constant.  
Here, the SU(2) currents 
and primaries are written in two bosons and Klein factors
\begin{eqnarray}
&&J_a ^+ (z)= \eta_{a \uparrow} \eta_{a \downarrow}
		e^{i \sqrt{2} \varphi_a(z)}, \ \ \
	J_a ^z(z) = \frac{i}{\sqrt{2}} \partial \varphi_a(z), \\
&&\bar{J}_a ^+ (z)= \bar{\eta}_{a \uparrow} \bar{\eta}_{a \downarrow}
		e^{-i \sqrt{2} \bar{\varphi}_a(\bar{z})}, \ \ \
	\bar{J}_a ^z(\bar{z}) = \frac{-i}{\sqrt{2}} 
		\bar{\partial} \bar{\varphi}_a(\bar{z}), \nonumber \\
&&g_{a \uparrow \uparrow} = \eta_{a \uparrow} \bar{\eta}_{a \uparrow}
		e^{i(\varphi_a+\bar{\varphi}_a)/\sqrt{2}}, \ \ \
	g_{a \uparrow \downarrow} = \eta_{a \uparrow} 
		\bar{\eta}_{a \downarrow} 
		e^{i(\varphi_a-\bar{\varphi}_a)/\sqrt{2}}, \nonumber \\
&&g_{a \downarrow \uparrow} = \eta_{a \downarrow} 
		\bar{\eta}_{a \uparrow}
		e^{-i(\varphi_a-\bar{\varphi}_a)/\sqrt{2}}, \
	g_{a \downarrow \downarrow} = \eta_{a \downarrow} 
		\bar{\eta}_{a \downarrow}
		e^{-i(\varphi_a+\bar{\varphi}_a)/\sqrt{2}}, \nonumber
\end{eqnarray}
with $a=1, 2$.  The Klein factors obey anticommutation relation
$
\{\eta_{a \alpha}, \eta_{b \beta} \}
= 2 \delta_{a b} \delta_{\alpha \beta},
$
to satisfy the following correct operator product expansion
 for the SU(2) symmetry
\begin{eqnarray}
J^k _a(z) g_{b \alpha \beta}(w,\bar{w}) &\sim&
	\frac{\delta_{ab}/2}{z-w} 
	(^t \sigma^k g_a)_{\alpha \beta}(w, \bar{w}), \nonumber \\
\bar{J}^k _a(\bar{z}) g_{b \alpha \beta} (w, \bar{w}) &\sim& 
	- \frac{\delta_{ab}/2}{\bar{z}-\bar{w}} 
	(g_a  {^t\sigma^k})_{\alpha \beta} (w, \bar{w}).
\end{eqnarray}
The interaction operators
\begin{equation}
S_{int} =  \int \frac{d^2 z}{2 \pi} \sum_{i=1} ^5
 \lambda_i \phi_i(z,\bar{z}),
\end{equation}
can be represented in terms of two $SU(2)$ currents and
primary fields of the WZW model
\begin{eqnarray}
&&\phi_1(z, \bar{z})=
 {\bf J}_1(z) \cdot \bar{{\bf J}}_1(\bar{z})+
 {\bf J}_2(z) \cdot \bar{{\bf J}}_2(\bar{z}),
\nonumber \\
&&\phi_2(z, \bar{z})=
 {\bf J}_1(z) \cdot \bar{{\bf J}}_2(\bar{z}) +
 {\bf J}_2(z) \cdot \bar{{\bf J}}_1(\bar{z}), 
\nonumber \\
&&\phi_3(z, \bar{z}) = tr [ g_1 i(\partial - \bar{\partial}) g_2] ,
\nonumber \\
&&\phi_4(z,\bar{z})=tr g_1 i(\partial -\bar{\partial}) tr g_2 ,
\nonumber \\
&&\phi_5(z, \bar{z})= {\bf J}_1 (z) \cdot {\bf J}_2(z) +
 \bar{{\bf J}}_1 (\bar{z}) \cdot \bar{{\bf J}}_2(\bar{z}),
\end{eqnarray}
where $\lambda_2 = J_1/\pi$, $\lambda_3 = - J_1 M^2 /\pi$, 
$\lambda_4 = J_1 M^2 /(2 \pi)$, $\lambda_5 = J_1/(2 \pi)$ and
$\lambda_1$ is a certain negative constant. 
The initial coupling constant $ \lambda_1$ is roughly estimated
as the order $1/10$
in numerical analysis for single linear chain 
\cite{AGSZ}. 
 Note that the
the perturbations  $\phi_3(z,\bar{z})$ and $\phi_4(z,\bar{z})$ include  
operators with the
conformal dimension $(\frac{3}{2}, \frac{1}{2})$ and
$(\frac{1}{2}, \frac{3}{2})$, which break the Lorentz and 
the parity symmetry \cite{NGE,CHP}.
The original spin model has SU(2) symmetry, translational symmetry and
a permutation with a translation of the one chain
\begin{equation}
{\bf S}_{i} \rightarrow {\bf T}_{i+1}, \ \ \ \ 
{\bf T}_i  \rightarrow {\bf S}_i,
\label{sym}
\end{equation}
which forbids interactions with dimension $x < 2$. If this symmetry is
broken, the symmetry breaking can be measured by the following order 
parameter
\begin{equation}
{\bf S}_i \cdot ({\bf T}_i-{\bf T}_{i+1}) \sim
tr[(g_1 + g_1^{\dag}) \frac{^t{\bm \sigma}}{2} ] \cdot
tr[(g_2 + g_2^{\dag}) \frac{^t{\bm \sigma}}{2} ].
\label{order}
\end{equation}
This order parameter also changes sign under a translation of one
spin ${\bf T}_i \rightarrow {\bf T}_{i+1}$.
The operator product expansion 
$$
\phi_i(z,\bar{z}) \phi_j(0,0) \sim \sum_{k} \frac{C_{ijk}}{|z|^2} \phi_k(0,0),
$$
gives one loop renormalization group beta function
$$
-l \frac{d \lambda_k}{d l} = \frac{1}{2} \sum_{ij} C_{ijk} \lambda_i \lambda_j,
$$
which has the following practical form 
\begin{eqnarray}
&&l \frac{d \lambda_1}{dl} = \lambda_1^2- \lambda_3 \lambda_4-\lambda_4^2,
\nonumber \\
&& l \frac{d \lambda_2}{dl} = \lambda_2^2+ \lambda_3 \lambda_4 + \lambda_3 ^2, 
\nonumber \\
&&l \frac{d \lambda_3}{dl} = 
-\frac{1}{2} \lambda_1 \lambda_3
+ \frac{3}{2} \lambda_2 \lambda_3 + \lambda_2 \lambda_4, 
\nonumber \\
&& l \frac{d \lambda_4}{dl} = 
\lambda_1 \lambda_3 + \frac{3}{2} \lambda_1 \lambda_4 
- \frac{1}{2} \lambda_2 \lambda_4, 
\nonumber \\
&& l \frac{d \lambda_5}{dl} =
\frac{1}{2} \lambda_3 \lambda_4.	
\label{RG}
\end{eqnarray}
Note that the beta functions have a symmetry 
$\lambda_3 \rightarrow -\lambda_3$, $\lambda_4 \rightarrow -\lambda_4$,
which corresponds to a translation on one chain 
${\bf T}_i \rightarrow {\bf T}_{i+1}$ in the original spin model.

\section{Renormalization group flow}

\begin{figure}[ht]
  \epsfxsize=3.3 in\centerline{\epsffile{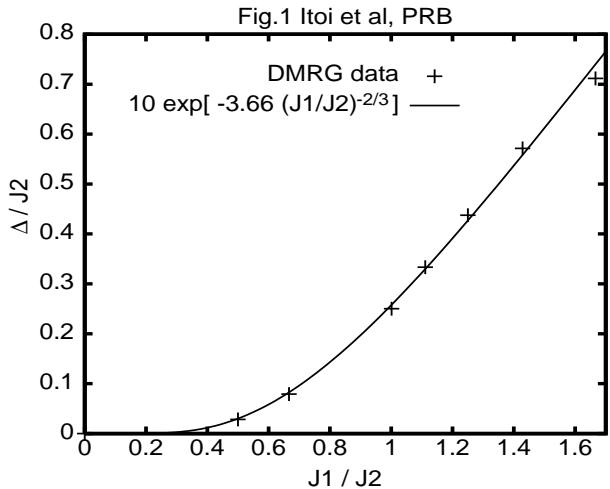}}
\vspace{0.5cm}
\caption[]{
The energy gap $\Delta$ vs. the antiferromagnetic interchain coupling 
$J_1 > 0 $ is depicted. The solid line is drawn by the scaling formula 
$\Delta \sim \exp(-c|J_1|^{-2/3})$
for small $J_1$ in the infinite order phase transition, 
using  Eq.(\ref{exp}) with $\xi\sim 1/\Delta$.
The gap $\Delta$ calculated by the DMRG
is obtained from Ref.[\onlinecite{WA}]. 
\label{figj1p}
}
\end{figure}

The critical point $J_1=0$ divides ferromagnetic and antiferromagnetic 
dimer phases.  Both regions have energy gap.
The renormalization flow 
diverges eventually along a stable direction 
$(\lambda_1, \lambda_2, \lambda_3, \lambda_4)
= (0,1, \frac{1}{\sqrt{2}},0)$ for the ferromagnetic coupling 
$J_1 < 0 $, and  along another stable direction 
$(\lambda_1, \lambda_2, \lambda_3, \lambda_4)=
(0,1, \frac{-1}{\sqrt{2}},0)$ for the antiferromagnetic coupling 
$J_1 > 0 $.  The current-current coupling $\lambda_2$ grows with 
positive value both in ferromagnetic and antiferromagnetic region.
In the effective field theory, the ferromagnetic model does not differ 
from the antiferromagnetic model transformed by
$
{\bf T}_i \rightarrow {\bf T}_{i+1},
$
which changes the sign of $\lambda_3$ and $\lambda_4$.  This fact 
suggests that both ferromagnetic and antiferromagnetic models have the 
same dimerization pattern. The correlation length $\xi$ behaves as
 \begin{equation}
 \xi \sim a \exp c |J_1|^{-\tilde{\nu}}
 \label{exp}
 \end{equation}
 with $\tilde{\nu}=2/3$ for $J_1 \sim 0$ and
the lattice spacing $a$. This result is obtained in another
RG method for the RG
equation \cite{IM2}.  In the 
antiferromagnetic region, this scaling formula of the correlation 
length holds for relatively large value in $0 < J_1 < J_c  
\sim 4.15 J_2$, even though
the formula is obtained merely by one loop approximation.  
We demonstrate it by numerical results from density matrix renormalization 
group calculation(DMRG) \cite{WA} in Fig. \ref{figj1p}.
The gap scaling formula can fit the numerical result quite well. 
This agreement even in relatively strong coupling region 
is not so surprising. There are many systems with Kosterlitz Thouless type
phase transition, where the gap scaling formula calculated in the weak coupling 
region holds even in the strong coupling region.
In the ferromagnetic region, however, this scaling formula 
does not hold for $|J_1| \sim J_2$. 
The energy gap in this region 
cannot be found in any numerical analysis. There should be some special
reason.  

We clarify in the following why the gap scaling formula in 
ferromagnetic region differs from the antiferromagnetic region.  The 
beta functions for $\lambda_1$, $\lambda_2$, $\lambda_3$, $\lambda_4$
have simultaneous zero on a line
\begin{equation}
\lambda_1=\lambda_2=0,
\  \lambda_3+\lambda_4=0.
\label{fixed}
\end{equation}
This line behaves like a fixed line for a four dimensional coupling 
constant space $(\lambda_1, \cdots, \lambda_4)$, since $\lambda_5$ does 
not enter their beta functions.  
Near this line, we define the deviation of the running coupling 
constants from the line by  $\lambda_3= \lambda + \delta \lambda_3$, 
$\lambda_4 = -\lambda + \delta \lambda_4$. The linearized 
beta functions are
\begin{eqnarray}
&&l \frac{d }{dl}(\delta \lambda_1 -\delta \lambda_2) \sim 0,
\nonumber \\
&& l \frac{d}{dl} (\delta \lambda_1 +\delta \lambda_2) \sim 2 \lambda
(\delta \lambda_3 + \delta \lambda_4), \nonumber \\
&&l \frac{d }{dl} (\delta \lambda_3 + \delta \lambda_4) \sim
-\lambda (\delta \lambda_1 - \delta \lambda_2), \nonumber \\
&& l \frac{d}{dl} (\delta \lambda_3 - \delta \lambda_4) \sim 0.
\end{eqnarray}
The eigenvalues of the 
scaling matrix all vanish on that line. This marginal property of the 
fixed line yields a remarkable phenomenon.  The linearized 
flow near the fixed line can be integrated as follows
\begin{eqnarray}
&&\delta \lambda_1(l) -\delta \lambda_2(l) \sim A, \nonumber \\
&&\delta \lambda_1(l) +\delta \lambda_2(l) \sim C + 2 \lambda B \ln l
		\nonumber \\
&&\delta \lambda_3(l) + \delta \lambda_4(l) \sim 
		B - \lambda A \ln l \nonumber \\
&&\delta \lambda_3(l) - \delta \lambda_4(l) \sim D,
\end{eqnarray}  
where $A, B, C$ and $D$ are integration constants determined by
initial coupling constants which  
are nonuniversal. If the initial coupling constants lie near 
this fixed line, $A+C$ is negative, $A-C \sim -2J_1/\pi$,
$B+D = -J_1 M^2/\pi-\lambda$ and $B-D=J_1 M^2/(2\pi) -\lambda$.
We can study the nature of the RG flow numerically
together with an analytic argument based on the behavior of the fixed line.  
Although the line is unstable except the case $A=B=0$, there is an 
extended region where the running couplings 
$\delta \lambda_1(l) +\delta \lambda_2(l)$ and 
$\delta \lambda_3(l) + \delta \lambda_4(l)$ flow toward $0$.  
In this case other couplings are renormalized logarithmically, and the 
running coupling constants spend a long time (long length scale) near 
the  fixed line (\ref{fixed}).  

\begin{figure}[ht]
\epsfxsize=3.3 in\centerline{\epsffile{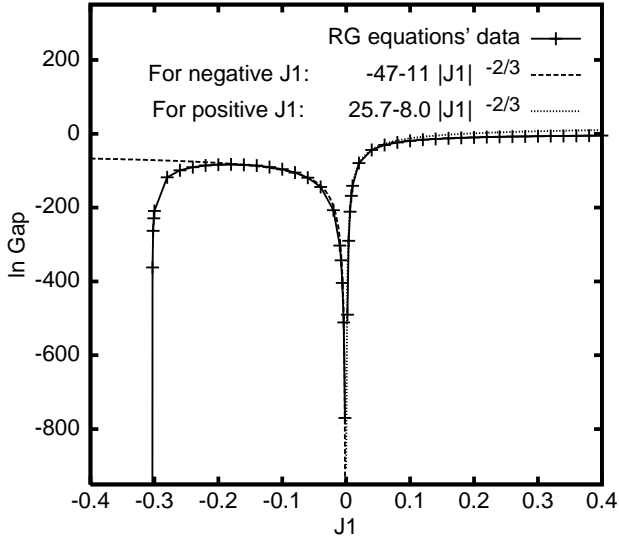}}
\vspace{0.5cm}
\caption[]{
The logarithm of gap $\ln \Delta$ vs. small $J_1$ estimated by
the scale where the solution of the RG equation has the singularity.  
The fitting line is obtained by fitting formula 
$\Delta= a \exp(-c/|J_1|^{2/3})$ with data close to zero.
Gaps obtained from RG equations in between $0.1 > J_1 > -0.1$ is 
used in fitting.
\label{fignegj1}
}
\end{figure}

In the ferromagnetic region with small $|J_1|$, we have 
$A < 0$, $B > 0$, $C < 0$, $\lambda > 0$. In this case,
$\delta \lambda_3(l) + \delta \lambda_4(l)$ does not flow toward 0, and
then the flow is free from the  fixed line as well as in the 
antiferromagnetic region.  In this 
region, we consider the gap scaling formula (\ref{exp}) holds and 
the gap is still small enough.
At $J_1 \sim -0.1 J_2 $ however, the 
constant $A$ changes sign and the running coupling constants start 
to flow toward 0. The correlation length grows again by the effect 
of the fixed line (\ref{fixed}).
The correlation length can be estimated approximately from the length 
scale $l$ where the running coupling constant $\lambda_i(l)$ diverges.  
The numerical solution of the RG equation shows a dependence of
the correlation length on the non-universal constant $M$. Typically, 
the minimal correlation length in the ferromagnetic region becomes an 
astronomical length scale more than $10^{36} a$ for $J_1 \sim -0.2 J_2$ 
and $M=1$. 
The interaction between two chains may change the non-universal quantities
$\lambda_1$ and $M^2$. 
However, the qualitative nature of the 
flow is unchanged for $M^2 < 2$ and $J_1 < 0$.
If $M^2 > 2.5$, the gap may be observed by numerical analysis.
We estimate the gap of the zigzag chain by numerically
evaluating the renormalization group equations.  
The data $\ln \Delta$ vs. $J_1$ is plotted in Fig. \ref{fignegj1}
setting
the nonuniversal quantities $M=1$ and 
$\lambda_1 = -0.24/\pi$ as typical values.
Also in this case we could observe asymmetric property 
between the ferromagnetic and antiferromagnetic coupling $J_1$.
The theoretical fitting function with $\tilde{\nu} =2/3$ 
can fit the data in the antiferromagnetic side $J_1 > 0$ 
better than those in the ferromagnetic side $J_1 < 0$.

This fact implies a quite unusual phenomenon. 
The infinite system differs essentially 
from a macroscopic system with the finite 
size. A system with a macroscopic finite size is described 
in a massless theory on the fixed line,  while an infinite system is 
described in a massive theory.  We can employ the massless field theory
for a macroscopic system available in an ordinary condensed matter 
experiment, since the correlation length
becomes an astronomical length scale. 
We can understand the quite slow convergence in 
numerical methods in this region. Although the beta functions depend 
on the renormalization scheme \cite{NGE,CHP}, the existence of the 
marginally relevant fixed line is scheme independent and the strong gap 
reduction in an extended region occurs universally.  On the other hand, 
in the antiferromagnetic interchain coupling, the flow is free 
from the  fixed line and the correlation length becomes $\xi \sim 7 a$ 
for $J_1 \sim J_2$.  This result is consistent with the numerical 
analysis in Ref.[\onlinecite{WA}] as depicted in Fig. \ref{figj1p}.

\section{Low energy levels}
Next we consider the effect of the chiral operators
$\phi_5(z)={\bf J}_1 (z) \cdot {\bf J}_2(z) $ and
$\bar{\phi}_5(\bar{z})=\bar{{\bf J}}_1 (\bar{z}) \cdot
\bar{{\bf J}}_2(\bar{z})$ with conformal dimension $(2, 0)$ and 
$(0, 2)$ respectively. In an ordinary macroscopic system size, we can 
employ the massless theory as a good effective theory.  A real fermion 
representation is useful to see the effect of the chiral operator 
\cite{AS,NGE,SNT}
\begin{eqnarray}
&&\xi_1(z) + i \xi_2 (z) = \sqrt{\frac{2}{a}} \eta_+
 \exp \left[-i \frac{\varphi_1(z)
  + \varphi_2(z)}{\sqrt{2}}\right] \nonumber \\
&&\xi_3(z) + i \xi_4(z) = \sqrt{\frac{2}{a}} \eta_-
 \exp\left[-i\frac{\varphi_1(z)-\varphi_2(z)}{\sqrt{2}}\right].
\end{eqnarray}
Here $\eta_{\pm}$ are written in a bilinear of ${\eta_{a \alpha}}$
such that the total current for the
SO(3) generator becomes
$
{\bf J}_1 + {\bf J}_2 = \frac{i}{2}  {\bm \xi} \times {\bm \xi},
$
where ${\bm \xi}=(\xi_1, \xi_2, \xi_3)$.  
Then the fields ${\bm \xi}$ and $\xi_4$ correspond to a triplet and
singlet excitation, respectively. In this mapping, the chiral 
perturbation operator has a free field representation
\begin{eqnarray}
\phi_5(z) &=& -\frac{1}{2} \partial \varphi_1 \partial \varphi_2 
	+ \eta_{1\uparrow} \eta_{1\downarrow} 
	\eta_{2\downarrow} \eta_{2\uparrow}
	\cos [\sqrt{2}(\varphi_1(z) - \varphi_2(z))] \nonumber \\
&=& \frac{1}{4} {\bm \xi} \cdot \partial {\bm \xi} 
	- \frac{3}{4} \xi_4 \partial \xi_4.
\label{chiraldeform}
\end{eqnarray}

The chiral perturbation operator (\ref{chiraldeform}) makes the shift 
of the spin wave velocities of triplet and singlet fields \cite{AS} 
\begin{equation}
v_t = v+ \frac{J_1}{4 \pi},  \ \ \ v_s = v- \frac{3 J_1}{4 \pi}.
\label{spvs}
\end{equation}
These shifts are found in the finite size correction to the low energy levels
of two triplet boundary operators $\xi_j \xi_k, \  \xi_i \xi_4$ with 
$j,k= 1,2,3$ in open boundary condition(OBC) and two triplet operators 
$\sigma_i \sigma_j \mu_k \mu_4$,  $\sigma_i \mu_j \mu_k \sigma_4$ for 
periodic boundary condition (PBC) \cite{AS}. Each gap in the leading 
finite size correction in OBC depends on $J_1$
\begin{eqnarray}
&&\Delta E_{jk} \sim \frac{\pi v_t}{L}=
 	\frac{\pi (v + J_1/{4 \pi} )}{L}, \nonumber \\
&&\Delta E_{j} \sim \frac{\pi(v_t+v_s)}{2L} =
 	\frac{\pi (v-J_1/{4 \pi})}{L}.
\end{eqnarray}
In PBC, all belong to the same energy independent of $J_1$
\begin{equation}
\Delta E \sim \frac{2\pi (3v_t+v_s)}{4L}=\frac{2 \pi v }{L}.
\label{pbcgap}
 \end{equation} 
There are some logarithmic corrections for a short chain mainly through
$\lambda_1(L)+\lambda_2(L) \sim 2/\ln L$ much larger than the effect of 
renormalization $\delta \lambda_5(L) \sim 6 \times 10^{-4} \ln L$.
We can expect the flat band instability at $J_1 =-4 \pi v$ in the 
fermionization.  This instability is a sign of a level crossing between 
the singlet and the ferromagnetic states.


\section{Numerical analysis}
These arguments are consistent with DMRG analysis on the OBC zigzag chain.
Here we chose $J_2=1$ and calculate the low energy levels numerically
by DMRG method\cite{whit}.  The gap is too small and the correlation 
length is too large to check the energy gap directly for $J_1 < 0$ in 
numerical analysis.  However, the results obtained in previous sections 
can be supported by the lowest three energy levels of the system. The 
ground state and first excited state for each chain at $J_1=0$ are 
unique and of total spin zero and one respectively \cite{egg} for even 
length chains.  We label the ground state for the zigzag chain as 
$(0,0)$ which is the product of the ground states of the two chains.  
The two degenerate first excited states of total spin one, $(0,1)^t$ 
and $(0,1)^s$, are the parity odd and even states as the product of the 
ground state and the first excited state from different chains.  The 
parity is for the symmetry of $S$ chain and $T$ chain permutation. 
 
\begin{figure}[ht]
\epsfxsize=3.3 in\centerline{\epsffile{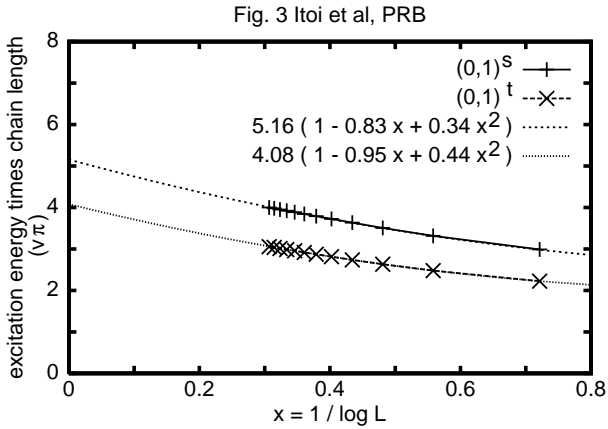}}
\vspace{0.5cm}
\caption[]{
The spin velocities for $J_1=-0.5$.  The excitation energies calculated
by DMRG for the state $(0,1)^t$ and $(0,1)^s$ multiplied by $L$ is 
plotted vs. $1/\ln L$ with $L$ up to 48.  Two different spin velocity 
$v_{t}$ and $v_{s}$ are obtained for them respectively by least square 
fitting.  The system behaves as its gap scaling being $1/L$.
\label{figlv}
}
\end{figure}

When $J_1$ decreases from zero, the excitation energy behaves as 
gapless for the infinite system in our finite size scaling fitting.  The 
energy of $(0,1)^t$, $\pi v_{t}/L$ decreases, and the one 
$\pi (v_{s}+v_{t})/2L$ of $(0,1)^s$ increases for small $|J_1|$. 
In Fig. \ref{figlv}, we demonstrate the two different spin velocities 
obtained by scaling for an arbitrary point $J_1=-0.5$ in the phase with
small and negative $J_1$.  We have kept $m=500$ states in DMRG 
calculation for OBC chains, and the biggest truncation error in DMRG is 
$10^{-7}$.

\begin{figure}[ht]
\epsfxsize=3.3 in\centerline{\epsffile{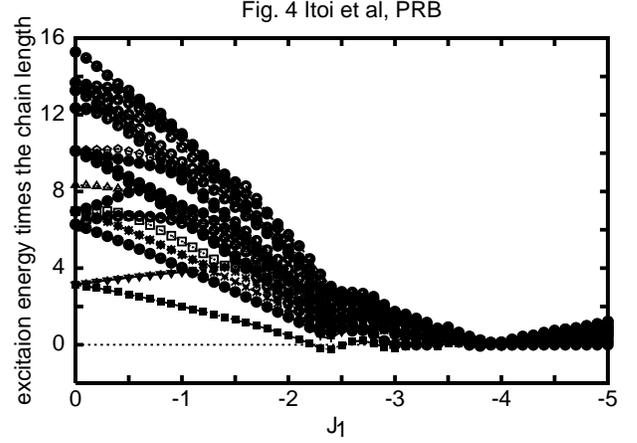}}
\vspace{0.5cm}
\caption[]{
Low excitation energies calculated by DMRG for zigzag open chain for
length $L=8$ with negative $J_1$.  
The lowest excited state at $J_1\sim 0$ is $(0,1)^t$ and the second
lowest is $(0,1)^s$.
The states collapse to zero 
energy when $J_1$ approaches $-4$. A non singlet ground state seems
to appear at $J_1 \sim -2$.

\label{figvst}
}
\end{figure}

The linear increasing of $v_{t}$ and decreasing of $v_{s}$ predicted in 
Eq.(\ref{spvs}) for $J_1\sim 0$ can be observed in Fig. \ref{figvst} 
by the low energy levels for OBC $L=8$ chains.  In Fig. \ref{figvst},
DMRG result is exact for such a short length $L=8$.  We have plotted 
the lowest five energy levels for each $S_z^{total}$ in the figure vs.
the interchain coupling $J_1$.

\begin{figure}[ht]
\epsfxsize=3.3 in\centerline{\epsffile{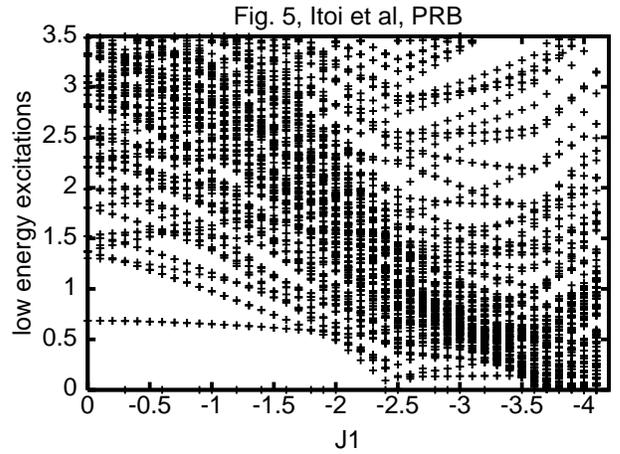}}
\vspace{0.5cm}
\caption[]{
Low excitation energies calculated by exact-diagonalization for 
zigzag PBC chains with $L=6$ (12 sites).  For each negative $J_1$,
eight lowest energy levels for each total momentum and $S_z^{total}$
is plotted.  The two lowest excited states at $J_1\sim 0$ are
of total spin 1.  Their energy is degenerate and almost independent
of $J_1$ as predicted in Eq.(\ref{pbcgap}).}
\label{figpbc}
\end{figure}

In Fig. \ref{figpbc}, the low energy levels obtained by exact 
diagonalization for a PBC $L=6$ chain is shown. We can check in the 
figure that the two lowest excitation energies are independent of $J_1$, 
as we expected in Eq.(\ref{pbcgap}).  The two lowest excitation 
energies are of total spin 1.  In the figure, eight lowest energy 
levels for each total momentum and $S_z^{total}$ of $L=6$ (12 sites) 
PBC chains has been plotted vs. interchain coupling $J_1$.

The low energy excitations are described in the critical theory near 
$J_1=0$ with the shifted spin wave velocities, therefore we can 
naturally exclude a different critical theory at a new fixed point 
away from $J_1=0$ which describes this system. 

In Fig. \ref{figvst} flat band instability is demonstrated. It occurs 
at $J_1=-4$ and the system enters the ferromagnetic phase with the 
maximum spin $S^{tot}=L$  ground state, which is consistent with 
previously obtained results \cite{HN,TH}. A non singlet ground state seems 
to occur near $J_1=-2$ as pointed by Cabra et al \cite{CHP}.  It is difficult to judge whether this 
partially polarized ground state can survive in the infinite length 
limit. In a macroscopic system size, it seems alive in $J_1 > -4$.

\section{discussions}
In conclusion, the divergent RG flow by the Lorentz symmetry
breaking perturbation certainly produces an energy gap 
in the zigzag chain with the
ferromagnetic interchain coupling as well as in the antiferromagnetic model.  
In an extended region of the ferromagnetic interchain coupling
with order 1, the correlation length can be extremely large 
due to the effect of the marginally relevant fixed line. 
If there is a finite energy gap, it becomes too tiny to observe.
The quantum fluctuation enhanced by the frustration 
yields the extraordinary reduction of the energy gap, 
which cannot be checked in ordinary macroscopic physics. 
The slow convergence in numerical analysis can be understood for this reason.
The DMRG analysis shows that 
the massless theory near the two decoupled chains
describes well the low energy levels in a macroscopic system.  

Here, we comment on an experimental result on the zigzag chain compound 
${\rm SrCuO_2}$ and the linear chain compound ${\rm Sr_2CuO_3}$. 
Motoyama, Eisaki and  Uchida show that the
temperature dependence of the spin susceptibility 
of the zigzag chain ${\rm SrCuO_2}$ differs from 
that in the linear chain ${\rm Sr_2CuO_3}$ \cite{MEU}. 
The drastic decreasing of the susceptibility is observed 
in the linear chain ${\rm Sr_2CuO_3}$ in low temperature.
This phenomenon is not observed in the 
zigzag chain ${\rm SrCuO_2}$, even though 
the three dimensional ordering temperature $\sim 2$ K of the zigzag
chain ${\rm SrCuO_2}$ is less than the temperature $\sim 5$ K of
the linear chain ${\rm Sr_2CuO_3}$. We would like to point out
the fact that the renormalization group flow
(\ref{RG}) has a certain parameter region of the ferromagnetic interchain
coupling in which
\begin{equation}
\chi(T) \sim \frac{1}{\pi v}\left(1-  
\lambda_1(1/T)/4-\lambda_2(1/T)/4 \right)
\end{equation}
does not decrease so drastically in the low temperature region.
It might be possible to understand the difference of the low temperature 
behavior of the susceptibility 
between the 
linear chain and the zigzag chain in terms of RG flow.

The universal scale reduction in the zigzag chain is 
interesting as a rare example from the view point of 
the hierarchy problem in the theoretical physics.  
A string theory as an ultimate unified theory 
of every thing in particle physics needs a natural 
explanation of the strong suppression of the energy scale.
Such a unified theory should explain the hierarchy 
of masses of the elementary excitations 
as well as the hierarchy of all interactions. We should obtain 
the mass of the light particle as a reduction 
from the Planck mass $\sim$ $10^{28}$ eV, for example,
the neutrino mass $\sim$ 1 eV, the electron mass $\sim 10^6$ eV, 
the muon mass $\sim 10^8$ eV, the Z boson mass $\sim 10^{11}$ eV. 
To understand this mass hierarchy problem in a unified theory, 
we have to answer the following two questions.
First, why do several energy scales 
appear depending on the species of the fundamental excitations ?
Second, how is the energy scale reduced with very large rate?
Normally, the small mass of fermionic excitations is considered as the 
result of a small chiral symmetry breaking. For the small mass
of bosonic excitations, the super symmetry \cite{S}
or composite approaches may work.
For the too large reduction rate, however, we have 
to tune the coupling constants to the corresponding small region. 
This is possible, but unnatural.
These two questions seem too difficult to resolve.
In condensed matter physics, however, we have interesting examples to
consider these problems.   
For the first question, we can refer the charge-spin separation in
the Tomonaga-Luttinger liquid.  Even though 
this model has the explicit symmetry breaking, 
the two fundamental excitations 
have different energy scales in the long distance physics.
For the second question, a stable large 
scale reduction occurs in the ferromagnetic zigzag chain model 
without any fine-tuning of the coupling constant $J_1$. 
In the almost models, as in the antiferromagnetic zigzag chain model,
however, the fine tuning of the  
coupling constant to the small region realizes 
the large scale reduction,
even if the spontaneous chiral symmetry breaking works there.
The experience on these examples may be useful
to find a solution of the mass hierarchy problem. 

This model should be studied still in 
more accurate approaches than the one loop renormalization group.
\\

The authors would like to thank I. Affleck for helpful discussions and 
informing the consistent result obtained by F. D. M. Haldane using a 
different approach.  They thank J. Pond for reading the manuscript 
and correcting
the presentation.
They are grateful to D. Allen for
correcting a mistake in the first version of this article.
Qin is partially supported by NSFC.

\end{document}